\documentstyle[preprint,prl,aps]{revtex}
\begin{document}
\draft
\title{Conductance Increase  by Electron-Phonon Interaction in Quantum Wires}
\author{Tobias Brandes, Arisato Kawabata}
\address{Department of Physics, Gakushuin University, 1--5--1 
Mejiro, Toshima--ku, Tokyo 171, Japan}
\maketitle
\vspace{0.5cm}
\begin{abstract}
We investigate the influence of electron-phonon interactions on the 
DC-conductance $\Gamma $ of a quantum wire in the limit of one occupied subband.
At zero temperature, a Tomonaga-Luttinger-like renormalization of $\Gamma $ to
a value slightly {\em larger} than $2e^{2}/h$ is calculated for a realistic 
quantum wire model. 
\end{abstract}
\section{Introduction}
In contrast to electron-electron scattering, electron-phonon (e-p) scattering 
in clean quantum wires at low temperatures is believed to be of minor 
importance for the DC-conductance change $\Delta \Gamma $ from the ballistic 
value $\Gamma =2e^{2}/h$. 
The extension of the Landauer-B\"uttiker formula \cite{Lan70,Bue86} to the 
interacting case is of theoretical interest in view of Fermi- and/or 
Luttinger-liquid behavior \cite{HDS93,Voi95} in quasi one-dimensional 
systems. 
Experimentally, $\Delta \Gamma \sim T^{\alpha }$ has been found recently 
\cite{THS95} with an exponent $\alpha $ in fair agreement with predictions 
from a Tomonaga-Luttinger model, while no consent was reached for the 
prefactor determining the $T=0$ value of the conductance. It was argued 
\cite{Pon95,Saf95} that the usual reduction of $\Gamma $ due to electron-
electron interactions \cite{AR82} is not observed in  realistic wires which 
are always coupled to leads where the interaction is screened. Thus, the 
electrons are free outside and interact only over a finite region 
within the wire which, upon using proper boundary conditions, finally
should reestablish the free value $\Gamma =2e^{2}/h$. 

However, in a recent paper \cite{Kaw96} it has been shown that 
this argumentation might not be complete. The reason is that the conductivity 
always has to be understood as the current response to the macroscopic 
(total) electric field and not to the external field (electric displacement), 
which in fact was pointed out by Izuyama \cite{Izu61} more than 30 years 
ago shortly after Kubo and Nakano had presented their theory of linear response. 
From the technical point of view, in an interacting system this difference in 
the fields exactly can be taken into account by {\em not} including 
'improper' diagrams simply connected by single Coulomb lines. Since the 
renormalization of $\Gamma $ in the Tomonaga- Luttinger approach \cite{AR82} 
is effectively equivalent to the RPA ('bubble-series' approximation)
\cite{Kaw96}, within this model 
the Coulomb interaction has no effect on the (properly defined) conductivity 
and hence the conductance itself which explains the absence of a 
possible deviation from the ballistic value observed in the experiment. 

Independent of the question if or if not Coulomb interactions are important 
for the conductance at low $T$, it is natural to ask which other processes 
can lead to deviations of $\Gamma $ from $2e^{2}/h$ in ballistic wires. 
Recently, interest has grown in the effect of electron-phonon coupling on 
Luttinger liquids \cite{MSG94,LM94,LM94a,BK94,Mar94,Kop95,HE95}. The 
electron-phonon scattering was shown to be a candidate for changes $\Delta 
\Gamma $ in previous works \cite{BK94,Mar94}, where at zero temperature the 
conductance turned out to be {\em increased}  by the e-p coupling in regular 
Luttinger liquids. In contrast to this, the DC-conductance of {\em chiral} 
Luttinger liquids, describing the edge of quantum Hall systems at filling 
factor $\nu$, turned out \cite{HE95} to be insensitive to phonons. 

In this paper, we present an alternative derivation of $\Delta \Gamma $ through 
perturbation theory in the e-p coupling and evaluate the conductance change 
for a realistic quantum wire at zero magnetic field. Summing up lowest order 
perturbation terms, we find a renormalization of $\Gamma$, 
\begin{equation} 
\label{gammarenorm}
\Gamma= \frac{2}{(1-\gamma )^{1/2}}\frac{e^{2}}{h}.
\end{equation} 
Note that since $\gamma >0$, the conductance is increased and not decreased, 
the e-p coupling thus acting like an effective attractive interaction between 
the electrons. We evaluate the numerical value of $\gamma $ for a wire 
embedded in a GaAs-AlGaAs heterostructure. Though $\gamma$ is of the order of 
$10^{-4}$ and the conductance change therefore very small, it at least 
demonstrates that the 'quantization' of the ballistic conductance is never 
complete in a realistic system. Furthermore, our calculation also shows that 
the comparatively simple perturbative result Eq.~(\ref{gammarenorm}) 
reproduces the calculation for the regular Luttinger liquid model 
\cite{BK94,Mar94}. This could indicate the limitations of the latter 
when applied to realistic quasi-1d systems, where for strong e-p coupling the 
perturbation theory breaks down. 

\section{Model}
In our model we start from the Hamiltonian 
\begin{eqnarray}
H&=&H_{0}+ H_{p}+ H_{ep}\nonumber\\
H_{0}&=&\sum_{\alpha }\varepsilon _{\alpha }c_{\alpha }^{+}c_{\alpha }\nonumber\\
H_{p}&=&\sum_{{\bf Q}}\omega _{Q}a^{+}_{{\bf Q}}a_{{\bf Q}}\nonumber\\
H_{ep}&=&\sum_{\alpha \beta {\bf Q}}M_{\alpha \beta }^{{\bf Q}}
c_{\alpha }^{+}c_{\beta }(a_{{\bf Q}}+a_{-{\bf Q}}^{+}),
\end{eqnarray}
where $\alpha $ refers to exact electronic eigenstates with energy 
$\varepsilon _{\alpha }$ of $H_{0}$, $a_{{\bf Q}}$, $a_{{\bf Q}}^{+}$ are
phonon annihilation and creation operators, and $M_{\alpha \beta }^{{\bf Q}}$ 
is the coupling matrix element.
To obtain transport quantities, Zubarev correlation functions
$
\langle\langle c_{\alpha }^{+}c_{\beta };c_{\gamma }^{+}c_{\delta 
}\rangle\rangle_{z}
$
are defined as
\begin{equation}
\label{zubarev}
\langle\langle A;B \rangle\rangle_{z}= -i\int_{0}^{\infty} dt e^{izt}\langle 
[A(t),B(0)]\rangle_{0},
\end{equation}
where the expectation value $\langle \phantom{A} \rangle_{0}$ refers to the 
equilibium density operator $\exp(-\beta H)/Z$, $\beta $ is the inverse 
temperature, and $Z=Tr\exp(-\beta H)$.
The Zubarev functions can be determined by using the equation of motions
in $z$--space,
$                                                 
z \langle\langle A;B \rangle\rangle_{z} -  
\langle\langle[A,H];B\rangle\rangle_{z} = \langle[A,B]\rangle_{0}.
$
To second order in the coupling,
the result consists of a term due to backscattering of electrons
with a momentum transfer $\approx 2k_{F}$, and a forward scattering term.
For real processes, in the limit of $T\to 0$, the backscattering term freezes 
out, and only the forward scattering survives. We do not consider virtual 
backward scattering processes \cite{Fukuyama} which corresponds to the 
absence of backscattering  ($g_{1}$)-processes in the Tomonaga-Luttinger 
model. 
The analytical expression
\begin{eqnarray}
\label{zubarevresult}
\langle\langle c_{\alpha }^{+}c_{\beta };c_{\gamma }^{+}c_{\delta 
}\rangle\rangle_{z}&=&
\delta _{\alpha \delta }\delta _{\beta \gamma }
\frac{f_{\alpha }-f_{\beta }}{z_{\alpha \beta }}+
\frac{1}{z_{\alpha \beta }}\frac{f_{\delta }-f_{\gamma }}{z_{\delta \gamma }}
\times\nonumber\\
&\times&\sum_{{\bf Q}}
\left[f_{\alpha }M_{\beta \alpha  }^{{\bf Q}} M_{\delta \gamma  }^{{\bf -Q}}
-f_{\beta }M_{\alpha \beta  }^{{\bf -Q}} M_{\gamma \delta  }^{{\bf Q}}
\right]
\left[\frac{1}{z-\omega _{ Q}}-\frac{1}{z+\omega _{ Q}}\right]
\end{eqnarray}
corresponds to the first term in a 'bubble' series in diagrammatic language.
Here, we introduced the abbreviations
$
z_{\alpha \beta }:=z+\varepsilon _{\alpha }-\varepsilon _{\beta }
$
and 
$
f_{\alpha }:= \left[\exp(\beta (\varepsilon _{\alpha }-\mu )) +1 
\right]^{-1}, 
$
where $\mu$ denotes the chemical potential in the Fermi distribution $f$.
The conductance is obtained
from the density-density correlation function
\begin{equation}
\label{densedense}
\chi(q,z) =i\int_{0}^{\infty}dt 
e^{izt}\langle[\rho_{q}(t),\rho_{-q}(0)]\rangle_{0},\quad \rho _{q}
:=\frac{1}{\sqrt{L_{s}}}\sum_{k\sigma }c^{+}_{k\sigma }c_{k-q\sigma },
\end{equation}
where the quantum number $k$ refers to plane waves in $x$-direction
\begin{equation}
\label{wavefunction}
\langle{\bf r}|k\rangle=\frac{1}{\sqrt{L_{s}}}e^{ikx}\phi(y)\chi(z),
\end{equation}
where $L_{s}$ is the length of the wire ($L_{s}\to \infty$ in the 
thermodynamic limit), $\phi(y)\chi(z)$ is the part of the wave function 
perpendicular to the wire, and $\sigma $ denotes the electron spin.
The relation of $\chi(q,z)$ to the conductivity $\sigma (q,z)$ is given 
through charge and current conservation,
$
\sigma(q,z)=-ie^{2}(z/q^{2})\chi(q,z).
$
With 
\begin{equation}
\chi(q,z)=
-\frac{1}{L_{s}}\sum_{kk'\sigma \sigma '}
 \langle\langle c_{k+q/2\sigma }^{+}c_{k-q/2\sigma  };
c_{k'-q/2\sigma '}^{+}c_{k'+q/2\sigma '}\rangle\rangle_{z},
\end{equation}
one obtains
\begin{eqnarray}
\label{chid}
\chi(q,z)&=&\chi_{0}(q,z)-\frac{1}{L_{s}}\sum_{kk'\sigma \sigma '}\frac{f_{k+q/2}-f_{k-
q/2}}{z+2kq}\frac{f_{k'+q/2}-f_{k'-q/2}}{z+2k'q}\times\nonumber\\
&\times&\sum_{{\bf Q}}M_{0,q}^{{\bf Q}} M_{0,-q}^{-{\bf Q}}
\frac{2\omega _{Q}}{z^{2}-\omega _{Q}^{2}}\nonumber\\
&=&\chi_{0}(q,z)- [\chi_{0}(q,z)]^{2}\frac{\pi  v_{F}}{2}\gamma _{q}(z),\quad
\gamma _{q}(z):=\frac{2L_{s}}{\pi \hbar^{2} v_{F}}
\sum _{{\bf Q}}|M_{k,k+q}^{{\bf Q}}|^{2}\frac{2\omega _{Q}}{z^{2}-
\omega _{Q}^{2}},
\end{eqnarray}

%
where in the definition of $\gamma _{q}(z)$ we reinstalled the $\hbar$.
Furthermore,
\begin{equation}
\chi_{0}(q,z)=-\frac{1}{L_{s}}\sum_{k\sigma }\frac{f_{k+q/2}-f_{k-q/2}}{z+2kq}=
-\frac{2v_{F}q^{2}}{\pi}\frac{1}{z^{2}-v_{F}^{2}q^{2}},\quad q\to 0, \quad k_{B}T 
\ll \mu
\end{equation}
is the density-density correlation function
for noninteracting electrons in one dimension including the spin. 
In the expression for $\gamma _{q}(z)$ we used the fact that the 
momentum matrix element $M_{kk'}^{{\bf Q}}$  depends 
on $k$ and $k'$ only through the difference $k-k'$ for 
plane waves so 
that $M_{k-q/2,k+q/2}^{{\bf Q}} M_{k'+q/2,k'-q/2}^{{-\bf Q}}=
|M_{k,k+q}^{{\bf Q}}|^{2}=|M_{0,q}^{{\bf Q}}|^{2}$.
Note that the expression  $\gamma _{q}(z)$ contains the phonon propagator
$\sim 1/(z^{2}-\omega _{Q}^{2})$ but no distribution functions and thus is
temperature independent.
Up to first order in $\gamma _{q}(z)$, one can write
\begin{equation}
\label{tomonagaexact}
\chi(q,z)=-\frac{2v_{F}q^{2}}{\pi}\frac{1}{z^{2}-v_{F}^{2}q^{2}(1+
\gamma _{q}(z))}.
\end{equation}
The appearance of the electron-phonon coupling term $\gamma _{q}(z)$ in the 
denominator in Eq.~(\ref{tomonagaexact}) suggests that the perturbative result
Eq.~(\ref{chid}) is the second term in a geometric series
\begin{equation}
\label{RPA}
\chi(q,z)= \chi_{0}(q,z)\sum_{n=0}^{\infty} \left[ -\chi_{0}(q,z) \frac{\pi 
v_{F}}{2} \gamma _{q}(z)\right]^{n}=
\frac{\chi_{0}(q,z)}{1+\chi_{0}(q,z)\frac{\pi v_{F}}{2}\gamma _{q}(z)}.
\end{equation}
Indeed, Eq.~(\ref{RPA}) is the standard random-phase approximation for the
density-density correlation function, which is obtained diagrammatically by 
summing up the bubble diagrams. Moreover, Eq.~(\ref{tomonagaexact}) is the 
exact (non-perturbative) result in a model that starts from the beginning by 
describing the electronic system in terms of density operators, i.e. in the 
sense of a Tomonaga-Luttinger liquid description \cite{BK94,Mar94} with
a linearized dispersion relation $\varepsilon _{k}=v_{F}|k|$. 

The conductance $\Gamma(z) $ is defined 
\cite{FL81} as the spatial average 
of the conductivity in real space over an interval of the length $L$,
\begin{equation}
\label{gammadefinition}
\Gamma(z) =\langle \sigma (q,z) \rangle_{L},\quad
\langle A \rangle_{L} = \frac{1}{L^{2}}\int_{-L/2}^{L/2}\int_{-L/2}^{L/2}
dxdx'\int_{-\infty}^{\infty}\frac{dq}{2\pi}e^{iq(x-x')}A(q)
\end{equation}
with the notation $\langle A \rangle_{L}$
for a function $A(q)$. The DC-limit $z\to0$ is determined by the 
$q\to0$ -behavior of $\chi(q,z)$, Eq.~(\ref{tomonagaexact}).
Using the expansion
$
\left\langle (z+v_{k}q)^{-1} \right\rangle_{L}=-(i/|v_{k}|)\left[
1/2+izL/(6|v_{k}|)+O(z^{2})\right],
$
one finds the conductance $\Gamma _{0}(z)$ in the 
non-interacting case as
\begin{equation}
\Gamma _{0}(z):=\langle \sigma _{0}(q,z) \rangle_{L}
=2\frac{e^{2}}{h}\left(
1+\frac{1}{3}\frac{izL}{v_{F}}+O(z^{2})\right),
\end{equation}
which is the well-known result \cite{Lan70,Bue86,KM89MK89} for the conductance of a ballistic one-
dimensional channel. 
With the notation
\begin{equation}
\label{gamdefinition}
\gamma:=-
\gamma _{q=0}(z=0):=\frac{2L_{s}}{\pi \hbar^{2} v_{F}}
\sum _{{\bf Q}}|M_{k,k}^{{\bf Q}}|^{2}\frac{2}{\omega _{Q}},
\end{equation}
we obtain from Eq.~(\ref{tomonagaexact}) and Eq.~(\ref{gammadefinition}) the 
result Eq.~(\ref{gammarenorm}), which agrees with the calculation with a 
bosonized electronic Hamiltonian \cite{BK94}. We can compare to the result by 
Apel and Rice \cite{AR82}, $\Gamma=g 2e^{2}/h$, where the interaction parameter 
$g$ is larger than unity for attractive and smaller than unity for repulsive 
electron-electron interaction. At $T=0$, the coupling to the phonons thus 
leads to an  effectively attractive interaction between the electrons which 
{\em increases} the conductance. Here, we will not discuss the implications 
of the singularity \cite{LM94} for $\gamma =1$  for strong e-p coupling since 
this regime is not accessible within our perturbative calculation. On the 
contrary, one has to assure that $\gamma \ll 1$ for consistency. We now show 
that this is indeed the case for a realistic wire, taking into account the 
full three dimensional phonon system with its coupling to the one dimensional 
wire for the evaluation of the factor $\gamma $, Eq.~(\ref{gamdefinition}). 

\section{Electron-Phonon-Coupling} 
For the
electron-phonon scattering in GaAs heterolayers, apart from the deformation 
potential interaction, the piezoelectric interaction is known to be important
especially at low temperatures. We follow  
Price \cite{Pri82} who has shown how to incorporate both contributions and to 
take into account the important effect of the screening by the 2DEG as well.
The fundamental coupling parameters are $\Xi$, the deformation potential, 
and $eh_{14}$ for the piezoelectric coupling, where $h_{14}$ is the 
piezoelectric coupling constant as the only non-vanishing component of the 
piezoelectric tensor for GaAs (zinc-blende structure). For the three 
dimensional phonon vector ${\bf Q}$ we use the notation
${\bf Q}=({\bf Q}_{\|},Q_{z})$, ${\bf Q}_{\|}=(Q_{x},Q_{y})$, $Q=|{\bf Q}|$,
and $Q_{\|}=|{\bf Q}_{\|}|$.
The vector ${\bf Q}_{\|}$ lies in the $x-y$ plane of the 2DEG. With the 
longitudinal and transversal sound velocities denoted by $c_{L}$ and $c_{T}$, 
respectively,  the e-p potential $|V_{{\bf Q}}|^{2}$ in the matrix element 
$|M_{kk}^{{\bf Q}}|^{2}=|V_{{\bf Q}}|^{2}|\langle k|e^{i{\bf Q}{\bf r}}|
k\rangle|^{2}$ is given by
\begin{equation}
\label{V}
\frac{1}{\omega _{Q}}|V_{\bf Q}|^{2}=\frac{\Xi ^{2}\hbar}{2\rho 
c_{L}^{2}\Omega }\left \{1
+\frac{1}{Q^{2}} \left( \frac{eh_{14}}{\Xi } \right)^{2}\left[
A_{L}({\bf Q})+ \left(\frac{c_{L}}{c_{T}}\right)^{2} A_{T}({\bf Q})
\right]S(Q_{\|})\right\}.
\end{equation}
The phonon frequency 
$\omega _{{\bf Q}}$ is $c_{L}Q$ or $c_{P}Q$ for the longitudinal and 
transversal phonons, respectively.                       
Furthermore, $\Omega $ is the crystal volume with mass density $\rho $,
$A_{L}({\bf Q})=9Q_{z}^{2}Q_{\|}^{4}/(2Q^{6})$ and
$A_{L}({\bf Q})+2A_{T}({\bf 
Q})=(8Q_{z}^{2}+Q_{\|}^{2})Q_{\|}^{2}/(2Q^{4})$,
and $S(Q_{\|})$ is the screening factor due to the mobile electrons in the
$x-y$ plane. We have included the latter only for the piezoelectric
coupling which diverges $\sim 1/Q$ for small wave vectors, and not for the 
deformation potential coupling. The latter is regular for $Q\to 0$ and
screening effects are already taken into account in the value of the 
deformation potential $\Xi $ itself. 
The screening function $S(Q_{\|})$ is given by
\begin{equation}
S(Q_{\|})=\frac{Q_{\|}}{Q_{\|}+PH( Q_{\|})},\quad
H(Q_{\|})=\int\int dz dz' \chi(z')^{2} \chi(z+z')^{2}\exp(-Q_{\|}|z|),
\end{equation}
where $H( Q_{\|})$ depends on the quantum well wave function $ \chi(z)$
and $P$ is the screening constant \cite{Pri82}. 
The matrix element
$|\langle k|e^{i{\bf Q}{\bf r}}|k\rangle|^{2}$
still contains the free electron wave function Eq.~(\ref{wavefunction})
of which the plane wave component in $x$ direction gives rise to a Kronecker 
$\delta _{Q_{x},0}$. This means that only phonons perpendicular to 
the wire contribute. On the other hand, the 'formfactor' due to the part of 
the wave function in direction perpendicular to the wire gives rise to a 
cutoff of phonons with wave vectors $Q_{z}\geq l_{z}^{-1}$ and $Q_{y}\geq 
l_{y}^{-1}$. 
Here, $l_{z}$ and $l_{y}$ are the thickness of the quantum layer and the 
quantum wire, respectively. 
We use a parabolic quantum well and a parabolic quantum wire 
confinement potential such that 
$|\langle k|e^{i{\bf Q}{\bf r}}|
k\rangle|^{2}=\delta _{Q_{x},0}\exp(-l_{z}^{2}Q_{z}^{2}-l_{y}^{2}Q_{y}^{2})$.
The evaluation of the renormalization factor Eq.~(\ref{gamdefinition})
is performed by introducing polar coordinates in the $y-z$ plane,
$Q_{z}=q\sin\varphi$, $Q_{y}=q\cos\varphi$, where $q=(Q_{y}^{2}+Q_{z}^{2})
^{1/2}$. Notice that the sum Eq.~(\ref{gamdefinition}) is two-dimensional 
because of the Kronecker $\delta _{Q_{x},0}$, furthermore the factor 
$L_{s}/\Omega $ is the area of the $y-z$ plane so that the transformation 
from summation to integration can be performed properly. 
We assume the thickness of the quantum well to be very small
compared to the wire width,
$l_{z}\ll l_{y}$, which allows us to evaluate the integrals analytically.
The result is
\begin{equation}
\gamma =\frac{2\Xi ^{2}}{(2\pi)^{2}\hbar v_{F}c_{L}^{2}\rho}
\left\{\frac{1}{l_{y}l_{z}}+
\left[\frac{9}{16}+\left(\frac{c_{L}}{c_{T}}\right)^{2}
\frac{13}{32}\right]g(Pl_{y})
\left( \frac{eh_{14}}{\Xi } \right)^{2}\right\},\quad
g(y):=\int_{0}^{\infty}dx\frac{e^{-x^{2}y^{2}}}{x+H(Px)}.
\end{equation}
The e-p parameters we used to obtain the numerical factor for $\gamma $ (table) are 
standard values for GaAs. It turns out that the contribution from the 
deformation potential coupling is of the same order as the  
piezoelectric contribution. 
Since $Pl_{y}\gg 1$, the value of $g(y)\approx \sqrt{\pi}/2y$ is due to $x\approx 0$ and by
$H(0)=1$ therefore nearly independent of the exact form of the quantum well 
wave function. We obtain
\begin{equation}
\label{numericalvalue}
\gamma \approx 3\cdot 10^{-4}\times \frac{1}{v_{F}[10^{4}ms^{-1}]}.
\end{equation}

\section{Conclusion}
The result $\gamma \ll 1$ is consistent with our perturbative approach. Note, 
however, that $\gamma \sim 1/v_{F}$ and therefore can in principle  be made 
arbitrarily large by tuning the Fermi velocity to very small values (near the 
bottom of the subband), corresponding to very high density of states. 
This region, however, can never be achieved for realistic values of $v_{F}$. 
The case $\gamma \approx 1$ where the theory breaks down corresponds to 
such small values as $v_{F}\approx 3 ms^{-1}$, where the conductance plateau 
in any case does not exist any longer. Still, from the technical point of 
view, the divergence of $\Gamma $ in this case indicates that in principle 
other higher order perturbation terms have to be considered. 
On the other hand, for Fermi energies well above the 
subband edge, our perturbation theory works well, $\gamma $ is small and 
leads to the increase of the conductance, Eq.~(\ref{gammarenorm}). The 
question remains if this increase really is observable in an experiment. 
Realistic quantum wires are never completely ballistic due to boundary 
roughness and other impurity effects. These, however, should in any case lead 
to a reduction and not to an increase of the conduction. Furthermore, the 
electron-electron interaction which originally \cite{AR82} was predicted to 
reduce $\Gamma$, turned out \cite{Pon95,Saf95,Kaw96} to be irrelevant for the 
$T=0$ value of $\Gamma $. The electron-phonon interaction, on the other hand, 
neither gives rise to the difference between macroscopic and external 
electric field in the sence of \cite{Kaw96,Izu61}, nor is it screened in the 
leads and locally unscreened in the wire as is the electron-electron 
interaction according to Ref. \cite{Pon95,Saf95}. It therefore should remain 
the only interaction mechanism to give a contribution to $\Delta \Gamma $ in 
the limit of zero temperature. From this point of view, we believe that it is 
worthwhile to investigate the prediction $\Gamma >2e^{2}/h$ in some more 
detail experimentally. 

One of the authors (T.B.) would like to acknowledge support by the EU STF9 
fellowship program in Japan.


\begin{table}[]
\label{parameters}
\caption[]{Electron-phonon parameters for evaluating $\gamma $ in GaAs. The first five 
parameters are taken from Ref. \cite{BFS93}}
\begin{tabular}{lcr}
Parameter&Symbol&Value\\
\tableline
Mass density&$\rho$& 5300 kg m$^{-3}$\\
Longitudinal speed of sound&$c_{L}$&5200 m s$^{-1}$\\
Transversal speed of sound&$c_{T}$&3000 m s$^{-1}$\\
Deformation Potential&$\Xi $&$2.2\times 10^{-18}$J\\
Piezoelectric constant&$eh_{14}$&$1.38\times 10^{9}eV$ m$^{-1}$\\
Screening constant\cite{Pri82}&$P$&$2\times 10^{8}$m$^{-1}$\\
Fermi velocity&$v_{F}$&$10^{4}$m s$^{-1}$\\
Quantum well width&$l_{z}$&$10^{-8}$m\\
Quantum wire width&$l_{y}$&$10^{-6}$m\\
\end{tabular}
\end{table}

\end{document}